\documentclass[letterpaper, 10 pt, conference]{ieeeconf}

\usepackage{authblk}
\usepackage{amsmath}
\usepackage{amssymb}
\usepackage{braket}
\usepackage{graphicx}
\usepackage{cite}
\usepackage{color}
\usepackage{soul}
\usepackage{pdfpages}

\title{\LARGE \bf
Continuous probe of cold complex molecules with infrared frequency comb spectroscopy
}

\author[1]{Ben Spaun}
\author[1]{P. Bryan Changala}
\author[2]{David Patterson}
\author[1]{Bryce J. Bjork}
\author[1]{Oliver H. Heckl}
\author[2]{ \authorcr John M. Doyle}
\author[1]{Jun Ye}
\affil[1]{JILA, National Institute of Standards and Technology and University of Colorado,  Department of Physics, \authorcr University of Colorado, Boulder, CO 80309, USA}
\affil[2]{Department of Physics, Harvard University, Cambridge, MA 02138, USA}

\begin{document}
\maketitle
\thispagestyle{empty}
\pagestyle{empty}

\begin{abstract}

Cavity-enhanced frequency comb spectroscopy for molecule detection in the mid-infrared powerfully combines high resolution, high sensitivity, and broad spectral coverage~\cite{Thorpe2008,Adler2010A,Foltynowicz2012}. However, this technique, and essentially all spectroscopic methods, is limited in application to relatively small, simple molecules. At room temperature, even molecules of modest size can occupy many millions of rovibrational states, resulting in highly congested spectra. Here we integrate comb spectroscopy with continuous, cold samples of molecules produced via buffer gas cooling~\cite{Patterson2010}, thus enabling the study of significantly more complex molecules. Until now, high resolution mid-IR spectroscopy of such molecules has been primarily performed with supersonic jets using single frequency lasers. In both sensitivity and spectral resolution, buffer gas sources are advantageous: they can be operated continuously, in contrast to most jets that are typically pulsed with a low duty cycle~\cite{Levy1980,Lovejoy1987}; and they provide samples moving slowly in the lab frame{, with no observation of buffer gas-molecule clustering}. Both of these features help to improve spectral resolution and detection sensitivity, as well as facilitate tracking of molecular interaction kinetics.  Here we report simultaneous gains in resolution, sensitivity, and bandwidth and demonstrate this combined capability with the first rotationally resolved direct absorption spectra in the CH stretch region of several complex molecules. These include nitromethane (CH$_3$NO$_2$), a model system that presents challenging questions to the understanding of large amplitude vibrational motion~\cite{Tannenbaum1956,Cox1972,Rohart1975,Sorensen1983,Sorensen1983a,Jones1968,McKean1976,Hill1991,Gorse1993,Hazra1994,Hazra1995,Cavagnat1997,Pal1997,Halonen1998,Dawadi2014}, as well as several large organic molecules with fundamental spectroscopic and astrochemical relevance, including naphthalene (C$_{10}$H$_8$)~\cite{SLS2011}, adamantane (C$_{10}$H$_{16}$)~\cite{Pirali2012}, and hexamethylenetetramine (C$_{6}$N$_4$H$_{12}$)~\cite{Bernstein1995,MunozCaro2004,Pirali2014,Pirali2015}. This general spectroscopic tool has the potential to significantly impact the field of molecular spectroscopy, simultaneously improving efficiency, spectral resolution, and specificity by orders of magnitude. This realization could open up new molecular species and new kinetics for precise investigations, including the study of complex molecules~\cite{Brumfield2012}, weakly bound clusters~\cite{Nesbitt2012}, and cold chemistry~\cite{Smith2006}.


\end{abstract}


For over half a century, high-resolution infrared spectroscopy has provided a broad foundation for the fundamental understanding of molecular structure and dynamics.  Spectroscopic analysis of vibrational, rotational, fine, and hyperfine structure yields extraordinarily detailed information about the molecular Hamiltonian. The success of such studies, however, requires well resolved and assignable spectra, which have been largely limited to relatively small, simple molecular systems. While the study of large and more complex molecules is an emerging front that provides fundamental insights into the nature of energy flow in a strongly correlated system, acquiring useful and tractable spectra for complicated molecular species poses many experimental challenges. Large molecules with many internal degrees of freedom have extremely large partition functions, with millions of quantum states occupied at room temperature. This leads to congested spectra with rotational structure unresolved in the presence of Doppler broadening. Moreover, floppy or non-rigid molecules have intrinsically complicated spectra with patterns that are difficult to interpret amid spectral congestion. A thorough understanding of more complex molecular systems requires the ability to acquire broadband spectra, spanning multiple vibrational bands, with both high resolution and high sensitivity.



Various infrared spectroscopic techniques have targeted one or more of these requirements. Supersonic expansions provide rotationally cold molecules~\cite{Lovejoy1987,Levy1980} and thus significantly reduce the congestion and linewidth of spectra, but typically at great cost to time-averaged molecule density. While low translational temperature is possible to achieve (such as along the slit direction in a slit jet expansion), molecules still have large lab-frame velocities that limit interaction and observation times. Single frequency continuous wave (cw) lasers coupled to optical enhancement cavities or multi-pass cells have achieved high resolution and sensitivity~\cite{Gagliardi2014}, but this approach lacks spectral bandwidth and therefore requires long acquisition times for broad spectral coverage. Conversely, Fourier transform infrared (FTIR) spectroscopy~\cite{Griffith2007} with broadband incoherent sources can quickly produce spectra spanning a wide frequency region, but this technique lacks sensitivity and is fundamentally limited in resolution by the optical path length of the spectrometer.

The recently developed technique of cavity-enhanced direct frequency comb spectroscopy (CE-DFCS) eliminates these technological limitations by combining the inherent broad bandwidth and high resolution of an optical frequency comb with the high detection sensitivity provided by a high-finesse enhancement cavity~\cite{Thorpe2006,Foltynowicz2013}. The broadband ($>$200~nm) spectrum of a frequency comb consists of tens of thousands of discrete, narrow frequency modes equally separated by the comb repetition rate, $f_\textrm{rep}$, with a common carrier envelope frequency offset, $f_\textrm{ceo}$. Both $f_\textrm{rep}$ and $f_\textrm{ceo}$ can be precisely stabilized, allowing for complete control of each of the tens of thousands of separate frequency modes in the comb~\cite{Udem1999,Diddams2000}. By matching the evenly spaced comb lines with the resonant frequency modes of a high finesse optical cavity filled with a molecular species, the absorption path length, and overall absorption sensitivity, of the comb can be enhanced by four orders of magnitude to kilometer length scales~\cite{Thorpe2006,Foltynowicz2011}. A Fourier-transform spectrometer (FTS) is used to measure the absorption spectrum and only needs a resolution better than the cavity free spectral range (100-1000~MHz) in order to resolve a single comb line. Thus, a standard FTS can achieve an instrument linewidth limited only by the stability of the comb itself, in our case $\sim$50~kHz. Providing high spectral brightness, frequency stability, and spatial coherence over a broad bandwidth, the massively parallel CE-DFCS technique is virtually equivalent to thousands of simultaneous, highly sensitive, absorption measurements with thousands of narrow linewidth lasers. The highly multiplexed nature of CE-DFCS has the potential to advance the field of infrared rovibrational spectroscopy just as the recent development of chirped-pulsed Fourier transform microwave spectroscopy has advanced the field of rotational spectroscopy \cite{Brown2008,Park2011}. Despite the potential of this spectroscopic technique, spectral congestion has thus far limited the application of CE-DFCS to small, relatively simple molecules with less than 10 atoms~\cite{Adler2010A, Adler2010B}.

\begin{figure}
\centering
\includegraphics[width=0.5\textwidth]{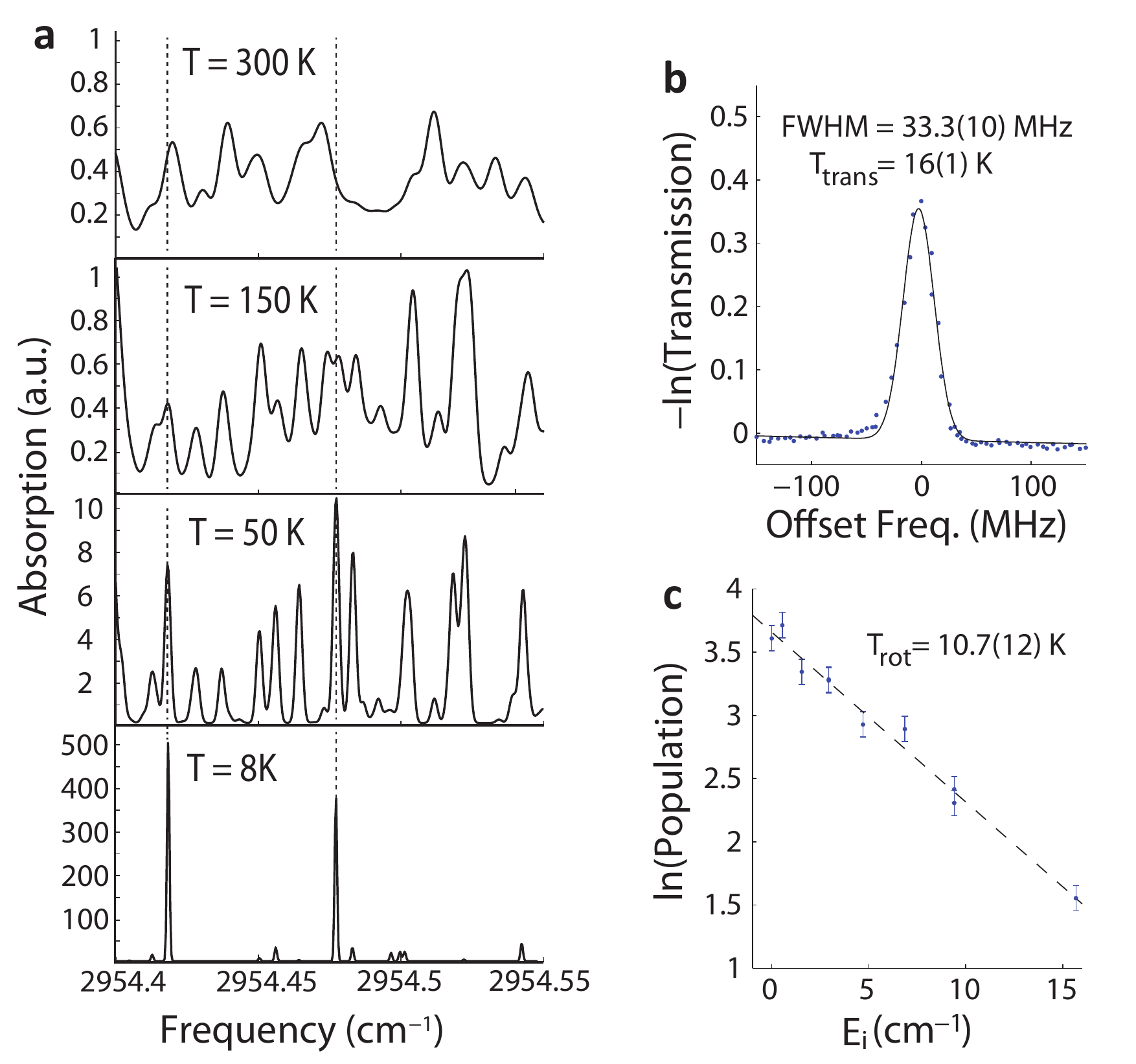}
\caption{\textbf{a} A simulated portion of the nitromethane spectrum as a function of temperature. Individual absorption lines are much more resolvable at low temperatures, as Doppler broadening decreases and the molecular population moves to the lowest available energy levels. \textbf{b} Observed Nitromethane Doppler-broadened absorption profile showing $\sim$16~K translational temperature. \textbf{c} Measured nitromethane rotational population as a function of energy, revealing a rotational temperature of $\sim$11~K (see Methods for more details)}
\label{fig:temperature}
\end{figure}

The work presented here applies CE-DFCS to significantly larger and more complex molecules by combining this spectroscopic technique with a buffer gas cooling method that can rotationally and translationally cool large molecules to $\sim$10~K and below~\cite{Patterson2012,Piskorski2015}. The simulated nitromethane (CH$_3$NO$_2$) absorption spectra in Fig.~\ref{fig:temperature}a show the significant gains in resolving power and sensitivity made possible by cooling molecules to $\sim$10~K. Colder molecules not only have a 5-times narrower Doppler-broadened linewidth, they also occupy many fewer, lower rovibrational energy levels. This results in a drastically simplified absorption spectrum, compared to the unresolvable room temperature spectrum \cite{Cavagnat1997}, with clearly distinguishable absorption lines with enhanced peak amplitudes. Without the requirements of high pumping speed and elaborate pumping infrastructure, buffer gas cells allow for higher resolution infrared molecular spectra with comparable absorption sensitivity, compared to state-of-the-art supersonic expansion jets~\cite{Davis1997, Brumfield2012, Patterson2012}. The $>$10~ms continuous molecular interrogation time provided by buffer gas systems is also orders of magnitude longer than the few-microsecond fly-through limited interrogation time of supersonic jets.~{We also note that we do not observe any evidence of complex formation between the cold molecules and buffer gas atoms (see Methods for details).} The combined buffer gas cooling and CE-DFCS apparatus allows us to quickly and easily resolve individual rovibrational transitions in complex molecules spanning multiple vibrational bands in the mid-IR using a simple molecular cooling apparatus.


\begin{figure} [!ht]
\centering
\includegraphics[width = 0.48\textwidth]{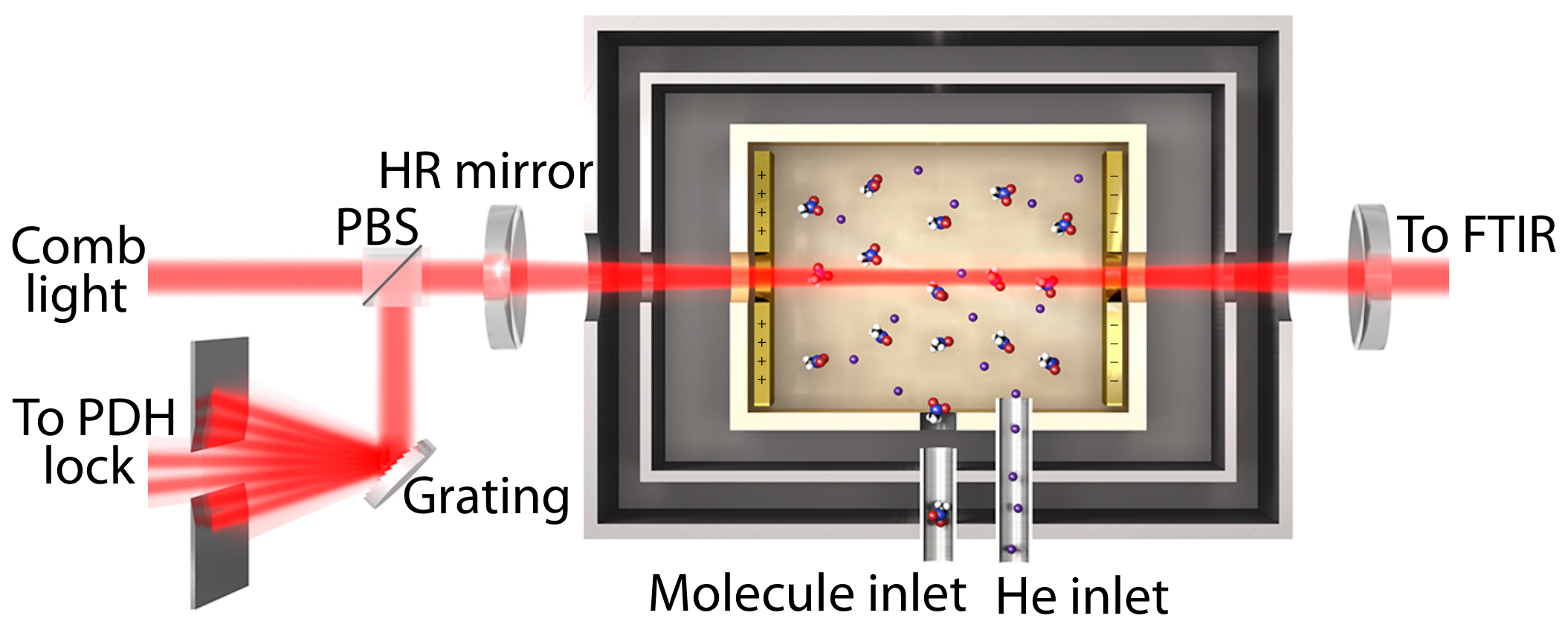}
\caption{A schematic of the combined CE-DFCS and buffer gas cooling apparatus. Light from a mid-IR frequency comb is coupled into a high-finesse enhancement cavity surrounding a 5-8~K buffer gas cell filled with cold molecules. Warm molecules enter through the side of the cell and are quickly cooled to $\sim$10~K through multiple collisions with the helium buffer gas. Comb light reflected from the cavity is dispersed by a grating, and a $\sim$10~nm segment of the comb spectrum is used to generate a Pound-Drever-Hall error signal to lock the comb to the cavity. The PDH lock allows for continuous transmission of thousands of frequency comb modes spanning $\sim$100~nm ~\cite{Thorpe2008,Adler2010A}. Transmitted comb light is coupled into a Fourier transform spectrometer, which measures the fractional absorption of each transmitted comb line.}
\label{fig:schematic}
\end{figure}


Figure~\ref{fig:schematic} shows the important components of the combined CE-DFCS and buffer gas cooling apparatus. A mid-IR frequency comb, tunable from 2.8~$\mu$m to 4.8~$\mu$m is produced in an optical parametric oscillator (OPO) pumped by a 1~$\mu$m ytterbium fiber comb~\cite{Adler2009, Foltynowicz2013}. Both $f_\textrm{rep} \approx 136$~MHz and $f_\textrm{ceo}$ of the mid-IR comb are referenced to a microwave cesium clock. The OPO comb output is then coupled into a high finesse ($F \approx 6000$) optical cavity surrounding a 5-8~K buffer gas cell. The cavity length is servoed via a piezo mirror actuator to ensure that the cavity free spectral range (FSR) is always exactly an integer multiple of $f_\textrm{rep}$. This allows many comb frequency modes over a broad bandwidth, limited by cavity mirror dispersion to $\sim$100~nm to be resonant with the optical cavity \cite{Foltynowicz2013}. Unlike white light sources, comb light is efficiently coupled into the enhancement cavity because the narrow linewidth of the comb is comparable to that of the cavity. The comb light makes thousands of round trips within the cavity, resulting in a 250 m total absorption path length with cold molecules in the buffer gas cell. To read out the fractional absorption of each comb mode, we use a home-built (doubled-passed) fast-scanning Fourier transform spectrometer (FTS) with a scanning arm (0.7~m) sufficiently long for single comb mode resolution~\cite{Adler2010B, Foltynowicz2013}. With the achievement of single comb mode resolution, our tabletop apparatus allows us to obtain broadband absorption spectra with an instrument linewidth of $\sim$50~kHz \cite{Adler2009}, more than two orders of magnitude narrower than the 20~MHz resolution of the best available white-light FTIR spectrometers, which use $\sim$10~m translation stages and are primarily available at user facilities \cite{SLS2011,SOLEIL2010}.

\begin{figure*}
\centering
\includegraphics[width=1\textwidth]{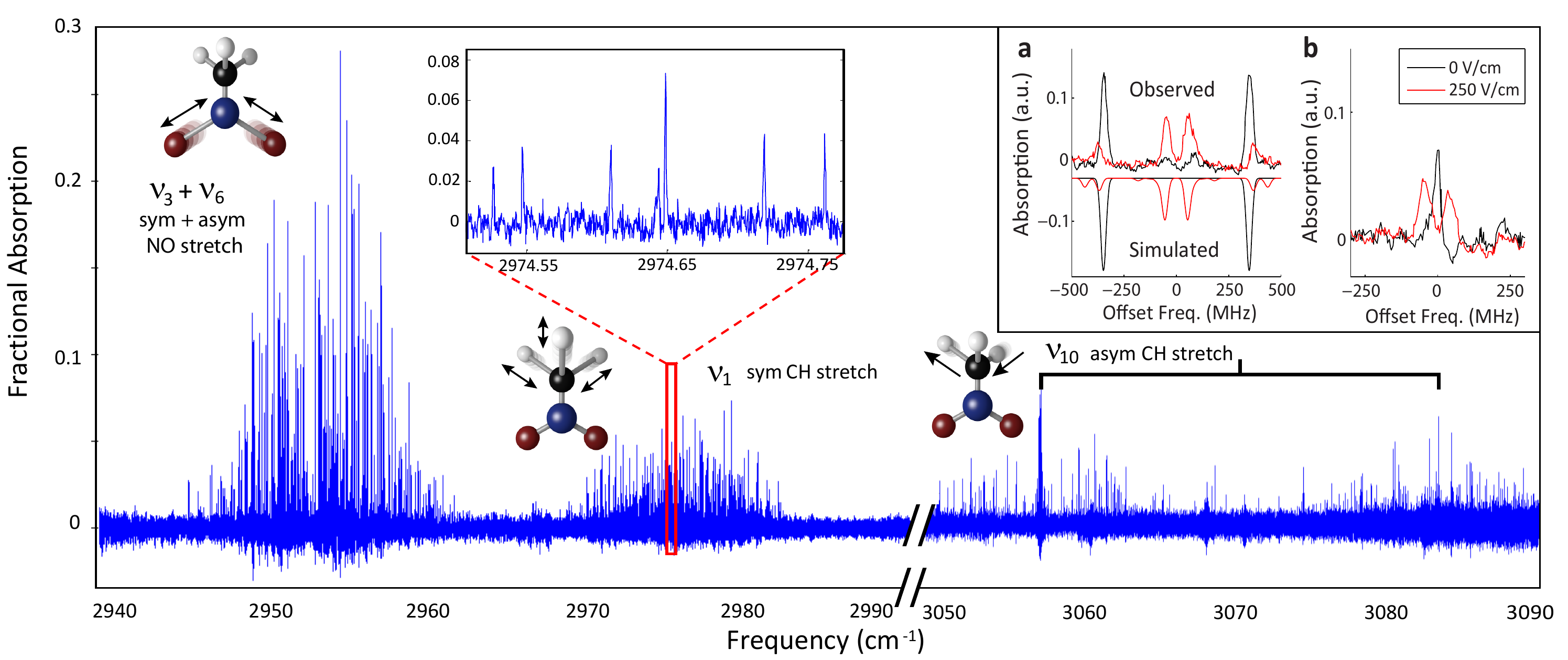}
\caption{Survey absorption spectrum of nitromethane, revealing over 1000 lines in multiple vibrational bands. The vibrational assignments are indicated in the figure, along with a description and illustration of the corresponding vibrational motions associated with each band. The left inset includes a small frequency window (0.2~cm$^{-1}$) of the 2974~cm$^{-1}$ band showing clearly resolved transitions and the typical spectral line density. Insets \textbf{a} and \textbf{b} show two examples of characteristic DC Stark splitting patterns, as described in the text.}
\label{fig:NitroMethSpectra}
\end{figure*}

The molecular absorption linewidth ($\Delta \nu$), dominated by 15-30~MHz Doppler broadening (Fig.~\ref{fig:temperature}b), is significantly smaller than the frequency spacing between comb modes transmitted through the enhancement cavity (272~MHz). Thus, for given values of $f_\textrm{rep}$ and $f_\textrm{ceo}$, the frequency comb is resonant with only a fraction of the molecular absorption features that lie within the comb bandwidth. To ensure that absorption lines are not missed by the discrete frequency comb modes, we step the comb repetition rate by $\Delta f_\textrm{rep}$ after averaging four FTS data acquisitions ($\sim$30~s total). This shifts the frequency of each comb mode by $n \times \Delta f_\textrm{rep}$, where $n \approx 10^6$ is the comb mode number. We choose $\Delta f_\textrm{rep}$ such that $n \times \Delta f_\textrm{rep} \lesssim \Delta \nu/5$, allowing us to measure the Doppler width, and therefore the translational temperature, of molecules in the buffer gas cell in real time. The complete spectrum containing all absorption lines is then generated by interleaving multiple FTS spectra, each corresponding to a different value of $f_\textrm{rep}$.


We demonstrate the simultaneous advantages in resolution, sensitivity, and bandwidth in this cold molecule-comb spectroscopy system by gathering rotationally resolved spectra in the CH stretch region ($\sim$3.3~$\mu$m) of nitromethane (CH$_3$NO$_2$), adamantane (C$_{10}$H$_{16}$), and hexamethylenetetramine (C$_{6}$N$_4$H$_{12}$) for the first time. CH$_3$NO$_2$ is a model system for understanding intramolecular vibrational coupling and large amplitude internal motion and has received considerable spectroscopic attention ~\cite{Tannenbaum1956,Cox1972,Rohart1975,Sorensen1983,Sorensen1983a}. 

Microwave studies have largely focused on the torsion-rotation structure of the ground vibrational state ~\cite{Tannenbaum1956,Cox1972,Rohart1975,Sorensen1983,Sorensen1983a}. There have also been several low resolution infrared experiments in the gas-phase~\cite{Jones1968,McKean1976,Hill1991,Gorse1993,Cavagnat1997} and, more recently, high resolution FTIR~\cite{Hazra1994,Hazra1995,Pal1997,Dawadi2014} and laser~\cite{Halonen1998} studies investigating the vibrational structure and the effects of internal rotation.

As shown in Fig.~\ref{fig:NitroMethSpectra}, we clearly resolve over 1000 nitromethane absorption lines spanning multiple vibrational bands, including the entire fundamental CH stretch region, with an excellent signal-to-noise ratio for less than three hours of data acquisition. The comb bandwidth is sufficiently large to simultaneously gather spectra containing the $\nu_3 + \nu_6$ (symmetric + antisymmetric) NO stretch combination band and the $\nu_1$ symmetric CH stretch band.\footnote{Our vibrational mode labeling convention follows that of Table 15-5 from Ref.~\cite{Bunker1998}.} The comb was also tuned to a higher center frequency to acquire the portion of the spectrum covering both components of the $\nu_{10}$ asymmetric CH stretch band.

Making use of existing nitromethane microwave data to provide ground state combination difference frequencies~\cite{Rohart1975,Sorensen1983}, we assigned transitions for several hundred mid-IR absorption lines, including those from excited torsional levels (see Methods for assignments, line lists, and rotational fits). The assigned rovibrational levels reveal interesting intramolecular rovibrational coupling at play in nitromethane.
For example, we observe energy level perturbations characteristic of anharmonic coupling between bright and dark rovibrational states. Most rotational term values of the $\nu_3 + \nu_6$ band fit well to a standard Watson A-reduced Hamiltonian. However, some levels are clearly perturbed by the nearby dark states of a separate vibrational level (see Fig. \ref{fig:DarkStates} in Methods for more detail). While coupling between the $\nu_3 + \nu_6$ level and higher quanta vibrational levels is possible, we suspect that the perturbing dark state is $\nu_5 + \nu_6$ ($\nu_5$ is the CH$_3$ umbrella bending mode), the only other two-quanta level expected in the $\sim$2950~cm$^{-1}$ region. In this case the relatively constant magnitude of the splitting between the mixed eigenstates, $\sim$0.14~cm$^{-1}$, suggests that the coupling between the methyl group vibration $\nu_5$ and the nitro group vibration $\nu_3$, which could manifest via a quartic $k_{3566}$ term in the anharmonic normal coordinate force field, is relatively weak. This is in stark contrast to the large zeroth-order splitting we observe between the in-plane and out-of-plane components of the nominally degenerate $\nu_{10}$ CH stretch band caused by interactions between the methyl and nitro groups. We also observe Coriolis coupling between the torsional, rotational, and vibrational angular momenta in the $\nu_3 + \nu_6$ and $\nu_{10}$ bands. The identification of lines and their spectral patterns is greatly simplified by the lack of systematic fluctuations in line intensities owing to the comb's capability of simultaneous acquisition of spectral features across the vast spectral region.

To simplify the line assignment process, we applied a moderate 50-400 V/cm tunable DC electric field within the buffer gas cell and monitored the distinct DC Stark-shift signature of each nitromethane absorption line. An electric field mixes molecular eigenstates together and causes them to experience a unique energy shift that depends on the molecular frame electric dipole moment and the presence of nearby states of opposite parity. As seen in the inset of Fig.~\ref{fig:NitroMethSpectra}, the high resolution and sensitivity provided by our apparatus allow us to clearly observe these relatively small ($\lesssim$100~MHz) energy level shifts in the nitromethane absorption spectrum. The pattern of the Stark shift is indicative of the specific types of eigenstates participating in the observed molecular transition. Figure~\ref{fig:NitroMethSpectra} inset b, for example, shows the clear Stark-splitting signature of a transition between excited torsional states ($|m| = 1$, where $m$ is the internal rotation quantum number), which split symmetrically and linearly in an electric field~\cite{Tannenbaum1956,Rohart1975}. Ground ($m=0$) torsional states, which are non-degenerate, exhibit no such first-order Stark splitting and can therefore be clearly distinguished from excited torsional states. Similarly, closely lying rotational levels of opposite parity will mix together, allowing new transitions to take place between pairs of mixed parity states, an effect clearly observed in Fig.~\ref{fig:NitroMethSpectra} inset a. The apparatus also helped resolve varying degrees of quadratic and linear energy shifts in many other molecular eigenstates when the modest electric field was applied. The information gained by comparing absorption spectra acquired with zero electric field to that acquired with a 50-400~V/cm electric field greatly facilitated the assignment process and provided additional confirmation of many line identifications.



\begin{figure*}
\centering
\includegraphics[width=1\textwidth]{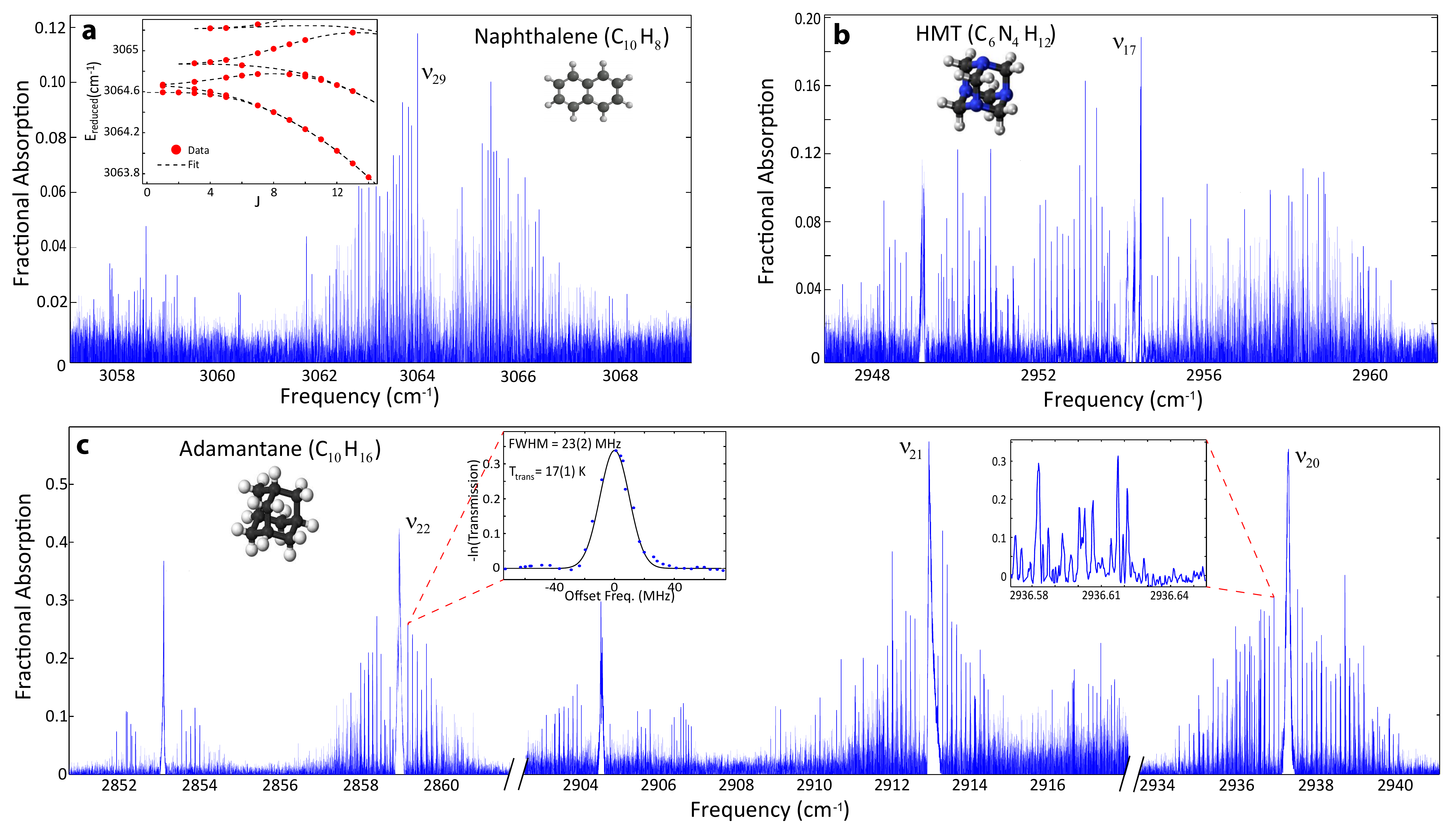}
\caption{Survey absorption spectrum of several large molecules including \textbf{a} naphthalene (C$_{10}$H$_{8}$), \textbf{b} hexamethylenetetramine (C$_6$H$_{12}$N$_4$), and \textbf{c} adamantane (C$_{10}$H$_{16}$) in the CH stretch fundamental region. In total, over 4000 absorption features were resolved in adamantane in 90~minutes of acquisition time, $\sim$1500 lines were resolved in hexamethylenetetramine in 30~minutes, and $\sim$1000 lines were resolved in naphthalene in 30~minutes. Known vibrational assignments are labeled, and insets in \textbf{c} reveal typical linespacing, Doppler-broadened linewidth, and detection noise floor. As illustrated by the inset of \textbf{a}, the observed reduced rotational term energies of the naphthalene $\nu_{29}$ band are well described by a semi-rigid asymmetric top effective Hamiltonian, with a residual scatter of 13~MHz.} 
\label{fig:LargeMol}
\end{figure*}

We have also used our apparatus to gather rotationally resolved spectra of several large organic molecules, shown in Fig. \ref{fig:LargeMol}, including naphthalene (C$_{10}$H$_8$), a polyaromatic hydrocarbon with an extensive spectroscopic history \cite{Pimentel1952,Hewett1994,SLS2011,Pirali2013}, adamantane (C$_{10}$H$_{16}$), the simplest member of the diamondoid family \cite{Pirali2012}, and hexamethylenetetramine (HMT, C$_6$N$_4$H$_{12}$), a molecule of astrochemical interest \cite{Bernstein1995,MunozCaro2004,Pirali2014,Pirali2015}. The high-resolution spectra of these molecules, which represent some of the largest molecules to ever be rotationally resolved in the mid-IR, were obtained with only 30-90~minutes of acquisition time per species.  As far as the authors are aware, these are the first rotationally resolved spectra to be obtained for these species in the CH stretch region \cite{Pirali2012,Pirali2014,Pirali2015}, with the exception of naphthalene for which skimmed-molecular beam optothermal experiments have been performed\cite{Hewett1994}, in contrast to the direct absorption spectra we report here. The adamantane spectrum (Fig. \ref{fig:LargeMol}c) reveals over 4000 absorption features spanning five separate vibrational bands comprised of the three IR active CH stretch fundamentals, $\nu_{20}$, $\nu_{21}$, $\nu_{22}$, and two other bands near 2853.1~cm$^{-1}$ and 2904.6~cm$^{-1}$. New unassigned bands are also observed in naphthalene and HMT near~3058 cm$^{-1}$ and 2954~cm$^{-1}$, respectively (Fig. \ref{fig:LargeMol}a-b). The inset in Fig.~\ref{fig:LargeMol}a shows the reduced term energies measured for rotational levels of the $\nu_{29}$~CH stretch band of naphthalene, illustrating very good agreement with the calculated effective Hamiltonian for a semi-rigid asymmetric top. Furthermore, we have measured the rotational and translational temperatures of these larger molecules and found them to be comparable to those of nitromethane, $\sim$10~K. (see Methods for additional analysis of large molecule spectra and line lists).

\begin{figure}
\centering
\includegraphics[width=0.45\textwidth]{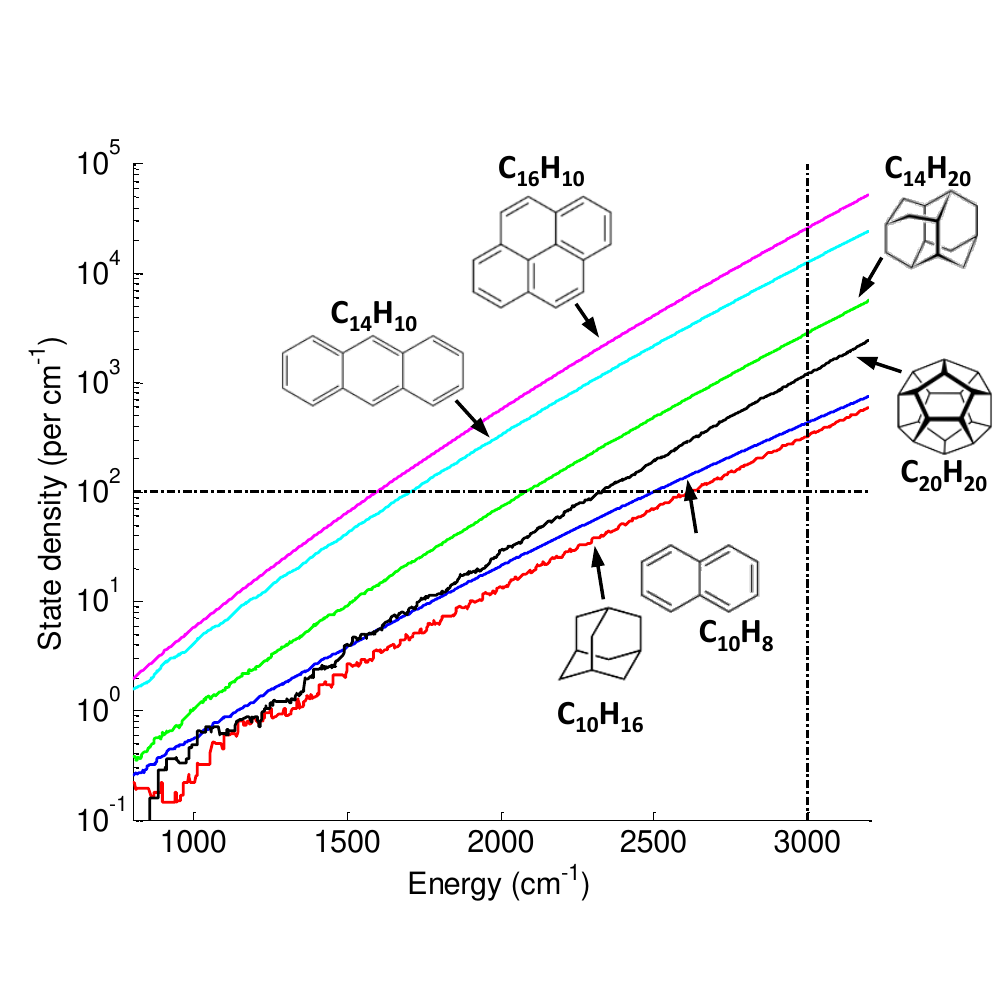}
\caption{The vibrational density of states for several large hydrocarbons. In increasing order, the total density of states (i.e. not symmetry selected) versus vibrational energy is shown for adamantane (C$_{10}$H$_{16}$), naphthalene (C$_{10}$H$_8$), dodecahedrane (C$_{20}$H$_{20}$), diamantane (C$_{14}$H$_{20}$), anthracene (C$_{14}$H$_{10}$), and pyrene (C$_{16}$H$_{10}$). These curves were calculated using a direct state count algorithm and a combination of previously observed and calculated vibrational frequencies (see Methods for details). The horizontal line at 100 states/cm$^{-1}$ marks the empirical threshold symmetry selected state density for IVR~\cite{Nesbitt1996,Buckingham2013}. The vertical line at 3000~cm$^{-1}$ indicates the approximate energy for CH stretch fundamental vibrations.}
\label{fig:DensityOfStates}
\end{figure}

\begin{figure}
\centering
\includegraphics[width = 0.45\textwidth]{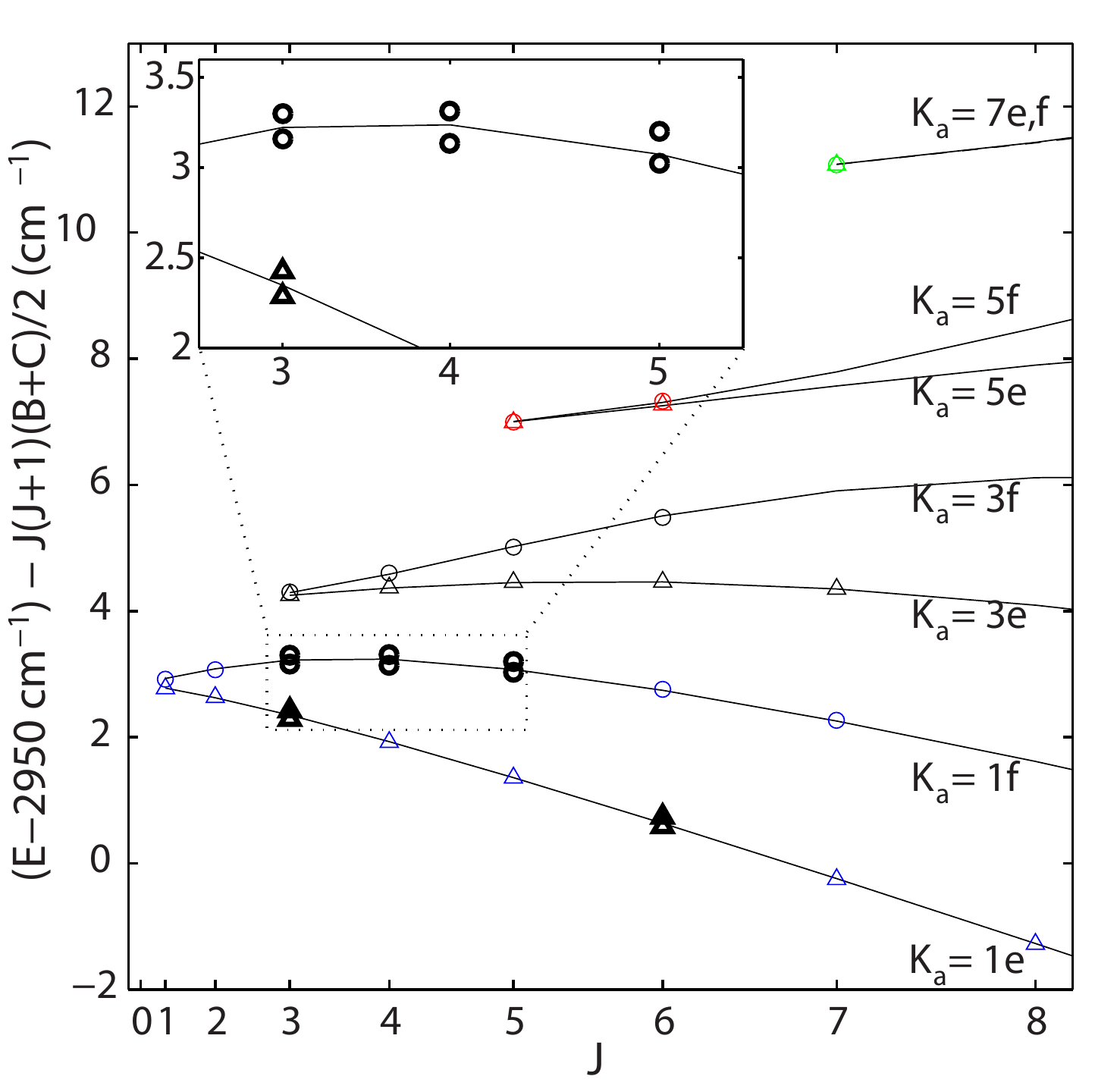}
\caption{Reduced term values of the rotational sub-levels of $\nu_3+\nu_6$ ($m=0$), plotted against the total angular momentum, $J$ (scaled as $J(J+1)$). The reduced energies are equal to the absolute energy minus $J(J+1)$ times the average of the $B$ and $C$ rotational constants. The solid lines connect sets of levels with respect to $K_a$ (the projection of $J$ onto the molecular inertial $a$-axis) and their parity (e/f) symmetry label. Inset: pairs of perturbed eigenstates, split symmetrically about the zeroth-order bright state position, are indicated in bold markers (see Methods for additional details).}
\label{fig:DarkStates}
\end{figure}

This novel cold molecule-comb spectroscopy system gives us the ability to quickly acquire rotationally resolved spectra of a number of large molecules in addition to those displayed in Fig. \ref{fig:LargeMol}. Intramolecular vibrational energy redistribution (IVR), however, presents an intrinsic challenge to high resolution spectroscopy of many large molecules in the 3-5~$\mu$m wavelength region of our frequency comb. As illustrated by Fig. \ref{fig:DensityOfStates}, the low vibrational state density of more rigid molecules, such as naphthalene and adamantane, prevents the onset of severe spectral fractionation due to anharmonic and rovibrational coupling between the observed bright states and the dense bath of dark states. On the other hand, we expect IVR to obscure the spectra of molecules with significantly higher state densities, such as pyrene or anthracene, in the CH stretch region. Rotationally resolved spectra of such molecules may only realistically be obtained at lower frequencies\cite{Brumfield2012}. As we push to acquire high resolution spectra of even larger molecules in the 3~$\mu$m region, highly symmetric, rigid species, such as dodecahedrane, C$_{20}$H$_{20}$, are the most promising targets. Many more molecules, such as C$_{60}$, will become accessible to this spectroscopic method as frequency comb technology is pushed deeper into the IR. 

The successful integration of buffer gas cooling of molecules and massively parallel direct frequency comb spectroscopy directly resulted in the first observation of thousands of fully resolved direct absorption features in the CH stretch region of nitromethane, naphthalene, adamantane, and hexamethylenetetramine. However, both the resolving power and detection sensitivity of this novel apparatus can still be improved, and its range of application expanded. For example, with simple modifications to the buffer gas cell we also foresee a straightforward path to the production and study of cold molecular radicals. The long ($>$10 ms) and continuous interrogation time provided by the buffer gas cell opens the door to many exciting new studies of the kinetics of cold radical reactions using our technique of time-resolved frequency comb spectroscopy \cite{Fleisher2014}. 

\vspace{3 mm}

\section*{Methods Summary}

Within the buffer gas cell, cold helium gas is used to cool gas-phase molecules to $\sim$10~K \cite{Patterson2012}. The 5-8~K aluminum cell (6~cm x 6~cm x 6~cm) is anchored to the cold stage of a pulse tube refrigerator and surrounded by a 35~K copper shield to minimize radiative blackbody heating. Electrically insulated flat copper electrodes on opposite sides of the cell allow for the application of tunable DC electric fields parallel to the cavity axis.  Between the cell and shield are helium cryopumps made of charcoal. These cryogenic components are enclosed within a  $\sim$$10^{-6}$~Torr vacuum chamber. Warm molecules enter the cell through a small $\sim$300~K tube, while a cold (5-8~K) tube delivers the buffer gas. The hot tube must be recessed 1-2~cm from the outer cell wall to prevent parasitic heating of the buffer gas. To achieve sufficiently high inlet flows of larger hydrocarbons, which are solid at room temperature, these species are first vaporized in a 50-200~$^\text{o}$C copper oven located just outside the 35~K blackbody shield. When the oven is sufficiently hot, a continuous flow of hydrocarbons exits the oven through a 2~mm aperture and then enters the cold cell. Molecules mix together in the cell where multiple cell-He and He-molecule collisions bring the initially warm molecules into thermal equilibrium with the cold cell. Measured molecular rotational and translational temperatures are typically $\sim$10-15~K (Fig.~\ref{fig:temperature}b-c), and molecular and helium densities are estimated to be $\sim$$5\times 10^{12}$~cm$^{-3}$ and $\sim$$10^{14}$~cm$^{-3}$, respectively. To study possible molecule clustering, higher polarizability neon is used as a buffer gas by adding thermal standoffs between the buffer gas cell and the refrigerator cold stage and warming the cell with heating resistors to $\sim$20~K.



\bibliography{library}
\bibliographystyle{natOutput}
\small
\textbf{Acknowledgements} We acknowledge funding from DARPA, AFOSR, NIST, and NSF-JILA PFC for this research. JD and DP acknowledge funding from the NSF and HQOC. BS is supported through the NRC Postdoctoral Fellowship. OHH is supported through the Humboldt Fellowship. PBC is supported by the NSF GRFP (Award No. DGE1144083). We thank Josh Baraban for many useful inputs and valuable discussions. We would like to thank Prof. David Perry for providing us with G. O. S{\o}rensen's original nitromethane ground state data.

\textbf{Author Contributions} All authors wrote this manuscript and contributed to the design, execution, and data analysis presented here. 

\textbf{Author Information} The authors declare no competing financial interests. Readers are welcome to comment on the online version of the paper. Correspondence and requests for materials should be addressed to BS (spaun@jila.harvard.edu) and JY (Ye@jila.colorado.edu).

\section*{\textbf{Methods}}

\textbf{Apparatus} A system of one-inch thick stainless steel rods and edge-welded bellows stabilizes the cavity length on a macroscopic scale and mechanically isolates the broadband (3.1-3.5~$\mu$m) high reflectivity cavity mirrors from the cold cell. The position and angle of each mirror is controlled by a set of three precision screws. One mirror is mounted to a tube-piezo for fine length adjustment with $\sim$1~kHz bandwidth.  The positions of the mirrors, precision mounts, and tube piezo are fixed on a macroscopic length scale by four 1-inch thick stainless steel rods which are  mechanically isolated from the vacuum apparatus. 

The length of the high finesse cavity are is servoed so that the cavity free spectral range (FSR) is always exactly twice the mid-IR comb repetition rate $f_\textrm{rep}$. Phase modulation for the Pound-Drever-Hall (PDH) error signal is
obtained by dithering the pump laser cavity length using a
fast PZT at one of its resonance frequencies (760~kHz). The
light reflected from the cavity is picked off with a polarizing beam splitter, dispersed with a reflection grating, passed through a slit, and directed on a photodiode. The grating and slit serve to select the comb spectral elements to which the cavity is locked. The photodiode signal from the $\sim$10~nm wide portion of comb light that passes through the slit is demodulated at the 760~kHz dither frequency to yield an error signal. This error signal is then used to a feed back on the cavity length, via the tube piezo, to fix the cavity FSR to $2 \times  f_\textrm{rep}$. Following the technique of \cite{Adler2009}, the mid-IR comb repetition rate ($f_\textrm{rep}$) and carrier envelope offset frequency ($f_\textrm{ceo}$) are each referenced to a frequency generated by a direct digital synthesizer (DDS) locked to a cesium clock \cite{Foltynowicz2013}.

\textbf{Fourier transform spectral processing} The path length difference $\Delta\ell$ of our Fourier transform interferometer is sufficiently long to ensure that the instrument linewidth, which is equal to $(\Delta\ell)^{-1}$, is smaller than the spacing between adjacent frequency comb modes transmitted by the enhancement cavity (i.e. the cavity free spectral range). By resolving individual comb modes, our spectral resolution becomes that of the frequency comb linewidth itself and is not limited by the instrument path length.

In order to exploit this drastic improvement in resolution, some post-processing must be applied to the acquired spectrum. The length of the interferogram we collect is typically such that the corresponding spacing between adjacent elements in the frequency domain is about 100~MHz. Since this is not an integer fraction of the absorption cavity FSR ($\sim 272$~MHz), the frequencies of the evenly spaced comb modes and the center frequencies of the Fourier transform spectrum walk on and off from each other. In order to measure the value of the spectrum at the actual frequencies of the comb modes, we resample the complex spectrum via convolution with the instrument lineshape function (a sinc function). This convolution can be performed efficiently, allowing us to easily and repeatedly resample the complex spectrum in order to locate the center frequency and intensity of each comb mode. Similar techniques, employing zero-padding of the interferogram, have been used recently for comb mode resolved Fourier transform spectroscopy by other workers as well~\cite{Maslowski2015}.

\renewcommand\thetable{\arabic{table}}
\begin{table}[ht]
\refstepcounter{table}\label{tab:dummy}
\end{table}

\textbf{Rotational fits: results and discussion}\\
\textbf{I. Nitromethane} A complete linelist of our assigned $m=$ 0 and 1 transitions of the 2953~cm$^{-1}$ band of CH$_3$NO$_2$, which we identify as $\nu_3 + \nu_6$, can be found in Table~\ref{tab:dummy}. Upper state term values were calculated using our measured transition frequencies and ground state torsion-rotation energies from previous studies~\cite{Rohart1975,Sorensen1983}. These energies were used to perform a least-squares fit of the $m=0$ levels, excluding perturbed states. Unfortunately, the number of $m=1$ assignments is too small to permit a fit of these levels. Watson's asymmetric top $A$-reduced Hamiltonian (I$^r$ representation)~\cite{Watson1977} was used as follows
\begin{align*}
\hat{H} &= A\hat{J}_z^2 + \frac{B+C}{2}\left( \hat{\mathbf{J}}^2 - \hat{J}_z^2\right) + \frac{B-C}{4}\left(\hat{J}_+^2 + \hat{J}_-^2\right)\\
&\quad -\Delta_J \hat{\mathbf{J}}^4 - \Delta_{JK} \hat{\mathbf{J}}^2\hat{J}_z^2 - \Delta_K \hat{J}_z^4\\
&\quad - \frac{1}{2}\left[\delta_J \hat{\mathbf{J}}^2 + \delta_K \hat{J}_z^2, \hat{J}_+^2 + \hat{J}_-^2\right]_+
\end{align*}
Given that the levels measured only extend to $J' \leq 8$, the quartic centrifugal distortion (CD) parameters were not well determined and therefore held fixed to their ground state values. Table~\ref{tab:fit} summarizes these results for the $\nu_3 + \nu_6$ band. It also includes the rotational parameters of the $m=0$ manifolds of the ground state and $\nu_6$ fundamental determined by previous studies~\cite{Rohart1975,Pal1997}.
\begin{table}[ht]
\caption{\label{tab:fit} {\normalfont Asymmetric top Hamiltonian fit results for CH$_3$NO$_2$. Values in brackets are fixed, and uncertainties are given in parentheses. All parameter values are specified in cm$^{-1}$. The inertial defect $\Delta_i$, determined from $I_c - I_a - I_b$, is given in u\AA$^2$.}}
\begin{tabular}{l|lll}
Parameter & Ground state$^a$ & $\nu_3 + \nu_6$$^b$&$\nu_6$$^c$\\
\hline
\hline
$T_v$ & 0 & 2952.6854(45)&1583.81163(20)\\
$A$ & 0.44503725(100)&0.439902(180)&0.4449620(33)\\
$B$ & 0.35172249(100)&0.347716(303)&0.3516825(26)\\
$C$ & 0.19599426(97)&0.194949(143)&0.1960255(9)\\
$\Delta_J \times 10^6$ & 0.2048(207)&[0.2048]&0.2431(23)\\
$\Delta_{JK} \times 10^6$ &0.5921(123)&[0.5921]&0.6822(103)\\
$\Delta_K \times 10^6$&-0.2515(133)&[-0.2515]&-1.5701(93)\\
$\delta_J \times 10^6$&0.08229(147)&[0.08229]&0.0717(11)\\
$\delta_K \times 10^6$&0.52536(634)&[0.52536]&0.4573(34)\\
RMSE&---&$1.1\times10^{-2}$&$2.3\times10^{-3}$\\
\hline
$\Delta_i$ & $+0.203$ & $-0.330$ & $+0.177$\\
\hline
\hline
\multicolumn{4}{l}{a. Ref.~\cite{Rohart1975}}\\
\multicolumn{4}{l}{b. This work}\\
\multicolumn{4}{l}{c. Ref.~\cite{Pal1997}}
\end{tabular}
\end{table}

The shifts of the rotational constants from the ground state to $\nu_3 + \nu_6$ are significantly larger in magnitude than those of the $\nu_6$ fundamental. Indeed, they appear larger than can be accounted for by excitation in $\nu_3$ alone, suggesting significant perturbation of this relatively highly excited level. Unfortunately, the $\nu_3$ fundamental has not yet been analyzed at high resolution, and so we cannot make a comparison of the rotational structure of both fundamentals and their combination band. Another way to quantify the change in the rotational structure is through the inertial defect ($\Delta_i = I_c - I_a - I_b$), which experiences a large negative shift from $+0.203$~u\AA$^2$ in the ground state to $-0.330$~u\AA$^2$ in $\nu_3 + \nu_6$. A negative change in the intertial defect is consistent with an increase in the torsional potential barrier~\cite{Halonen1998}, as the methyl group becomes locked-in with respect to the plane of the C-NO$_2$ frame. We note that this is also consistent with our very preliminary analysis of currently assigned $m=1$ levels. However, further progress on $m>0$ assignments is necessary before more definitive conclusions can be made.

\begin{table}[ht]
\refstepcounter{table}\label{tab:dummyNaph}
\end{table}
\textbf{II. Naphthalene}
The $\nu_{29}$ band of naphthalene was also treated using the asymmetric top effective Hamiltonian above. The subset of levels used in the fit ($J' \leq 14, K_a' \leq 3$), again, did not permit a determination of the quartic CD parameters, which we held fixed to ground state values from Ref.~\cite{Pirali2013}. A listing of the 155 $b$-type transitions we assigned and included in this fit is given in Table~\ref{tab:dummyNaph}. The fitted molecular constants are summarized in Table~\ref{tab:naphFit}.
\begin{table}[ht]
\caption{\label{tab:naphFit} {\normalfont Asymmetric top Hamiltonian fit results for naphthalene. Values in brackets are fixed, and uncertainties are given in parentheses.}}
\begin{tabular}{l|lll}
Parameter & Ground state$^a$ & $\nu_{29}$$^b$&$\nu_{29}$$^c$\\
\hline
\hline
$T_v$ & 0 & 3064.5942(5)&3064.58(2)\\
$A$ & 0.104051836(124)&0.104198(30)&0.104013(17)\\
$B$ & 0.04112733(37)&0.0411173(38)&0.0411023(45)\\
$C$ & 0.029483552(140)&0.02942455(9)&0.0294062(20)\\
$\Delta_J \times 10^9$ & 0.528(49)&[0.528]&\\
$\Delta_{JK} \times 10^9$ &1.206(145)&[1.206]&\\
$\Delta_K \times 10^9$&5.648(112)&[5.648]&\\
$\delta_J \times 10^9$&0.1752$^d$&[0.1752]&\\
$\delta_K \times 10^9$&1.951$^d$&[1.951]&\\
RMSE&3.1$\times10^{-4}$&4.3$\times10^{-4}$&$6.3\times10^{-4}$\\
\hline
$\Delta_i$ & $-0.137$ & $+1.137$ & $+1.057$\\
\hline
\hline
\multicolumn{4}{l}{a. Ref.~\cite{Pirali2013}}\\
\multicolumn{4}{l}{b. This work}\\
\multicolumn{4}{l}{c. Ref.~\cite{Hewett1994}}\\
\multicolumn{4}{l}{d. Calculated values~\cite{Pirali2013}.}
\end{tabular}
\end{table}
The values of the $A$, $B$, and $C$ rotational constants of the $\nu_{29}$ level are similar to the ground state values. The equilibrium geometry of naphthalene is planar, which is consistent with the small inertial defect of the ground state. However, the defect significantly increases upon excitation of the $\nu_{29}$ in-plane CH stretching mode. Rotational constants for this band have been previously measured in a skimmed molecule beam experiment~\cite{Hewett1994}, but differ significantly from the values reported here. Indeed, the ground state rotational constants from that study do not agree with our observed ground state combination differences, which were well reproduced using the values listed in Table~\ref{tab:naphFit} from Ref.~\cite{Pirali2013}.
\\

\textbf{Rotational temperature} In order to determine the rotational temperature of the buffer gas cooled nitromethane molecules, we first calculated the relative population in different rotational levels of the torsional-vibrational ground state. To do this, Hamiltonian fit parameters (see above) were used to construct rotational Hamiltonian matrices, which were diagonalized to produce rotational eigenfunctions for both the ground and $\nu_3 + \nu_6$ levels. Transition dipole matrix elements between these rotational eigenfunctions can be calculated using well-known direction cosine matrix elements~\cite{Townes1975}. The peak intensities of P and R-branch transitions with $J''=0-8$ and $K_a''=0$ (which we estimate to have a measurement uncertainty of 10\%) were normalized by the square of the corresponding transition dipole matrix elements to generate the relative populations in each level. The logarithm of these relative populations as a function of energy was then fit to a first-order polynomial (Fig. 1 in the main text), the slope of which is equal to $-(kT_{\text{rot}})^{-1}$, where $k$ is the Boltzmann constant and $T_\text{rot}$ is the effective rotational temperature. Our extracted rotational temperature of 10.7(12)~K is comparable to the translational temperature, 16(1)~K, determined by the measured Doppler widths. Thermalization between translational and rotational degrees of freedom is only partial, but certainly more complete than that typically obtainable in a supersonic jet.

\begin{figure}[ht]
\centering
\includegraphics[width=3.5in]{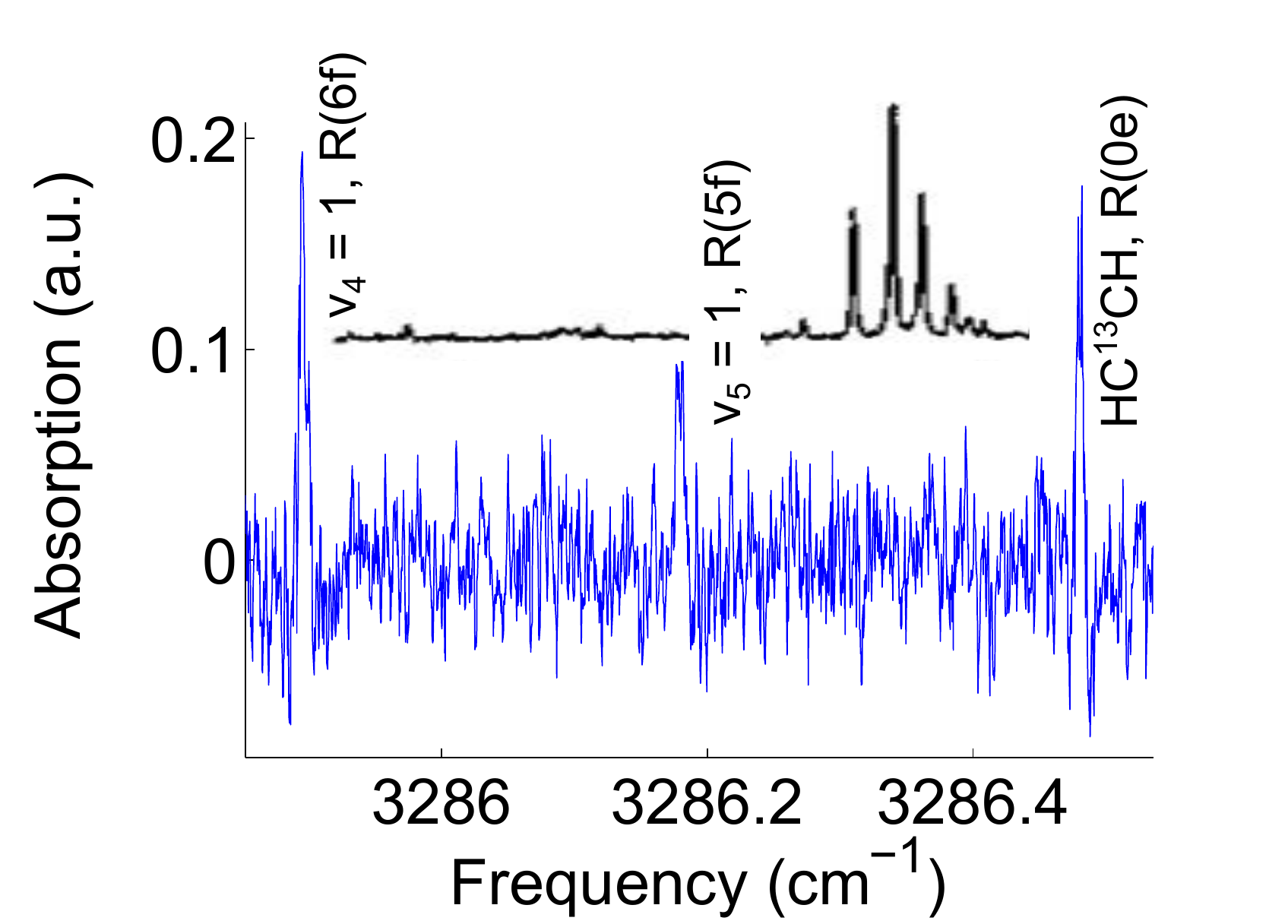}
\caption{\label{fig:nec2h2} Comparison of measured buffer gas cooled C$_2$H$_2$ spectrum (bottom trace) with that of the Ne-C$_2$H$_2$ complex (upper trace; reproduced from Fig. 1 of Ref.~\cite{Bemish1998}, permission pending). Three acetylene monomer transitions in the buffer gas cooled spectrum, including two hot band transitions and a $^{13}$C feature as described in the text, have been labeled. The buffer gas cooled spectrum has been rebinned with a bin size of 5 frequency elements ($\sim$40~MHz total).}
\end{figure}

\textbf{Molecule-buffer gas clustering} Because of the cold buffer gas temperature, the formation of weakly bound complexes containing a molecule and buffer gas atom(s) is possible. We attempted to observe neon-acetylene, Ne-C$_2$H$_2$, complexes by cooling C$_2$H$_2$ in a neon buffer gas at 20~K. In Fig.~\ref{fig:nec2h2}, we compare the previously measured spectrum of the complex (upper trace, reproduced from Ref.~\cite{Bemish1998}), with our measured spectrum of the buffer gas cell (lower trace). We observe no absorption from Ne-C$_2$H$_2$ above our baseline noise floor. The acetylene flow rate into the cell for this measurement (10~sccm) was sufficiently high to saturate our absorption dynamic range for most of the monomer transitions. Therefore, to aid in the comparison of the relative absorption of the complex, we have marked the R(0) transition at 3286.476~cm$^{-1}$ of the $\nu_3$ band of HC$^{13}$CH, which occurs at about 1\% natural abundance relative to the normal isotopologue. Two hot band transitions, at 3285.891 and 3286.176~cm$^{-1}$, from vibrational levels with one quantum of excitation in either of the two degenerate bending modes [$(\nu_2+2\nu_4+\nu_5)^{\ell=1}_{\text{II}}-\nu_4$, R(6f) and $(\nu_2+\nu_4+2\nu_5)^{\ell=1}_{\text{II}}-\nu_5$, R(5f)] are also labeled in the cold cell spectrum. Based on this measurement, we estimate the peak absorption of Ne-C$_2$H$_2$ to be less than 0.1\% relative to the monomer. This upper bound complements previous experiments that determined the population fraction of He-\textit{trans}-Stilbene complexes to be less than 5\% using UV-LIF spectroscopy in a similar cold cell apparatus~\cite{Piskorski2015}.

\textbf{Vibrational density of states} The vibrational density of states estimates presented in Fig.~\ref{fig:DensityOfStates} were calculated using a direct state counting algorithm~\cite{Baer1996}. We used observed--and when not available, calculated--vibrational frequencies for adamantane~\cite{Bistricic1995,Jensen2004}, naphthalene~\cite{Hewett1994,Sellers1985,Mitra1959}, diamantane~\cite{Ramachandran2006,Jenkins1980}, dodecahedrane~\cite{Hudson2005,Karpushenkava2009}, anthracene~\cite{Szczepanski1993,Bakke1979}, and pyrene~\cite{Vala1994,Shinohara1998}. In the case of only purely vibrational anharmonic interactions, the relevant density of states is that of states with the same vibrational symmetry as the zero-order bright state. This fraction is $n^2/g$, where $g$ is the order of the molecular point group and $n$ is the dimension of the irreducible representation of interest~\cite{Pechukas1984}. For the CH stretch fundamental levels, in particular, these fractions are 1/8 for naphthalene, anthracene, and pyrene; 3/8 for adamantane; 3/40 for dodecahedrane; and 1/12 or 1/3 for diamantane (non-degenerate or doubly degenerate modes, respectively).

\newpage
\includepdf[pages=-]{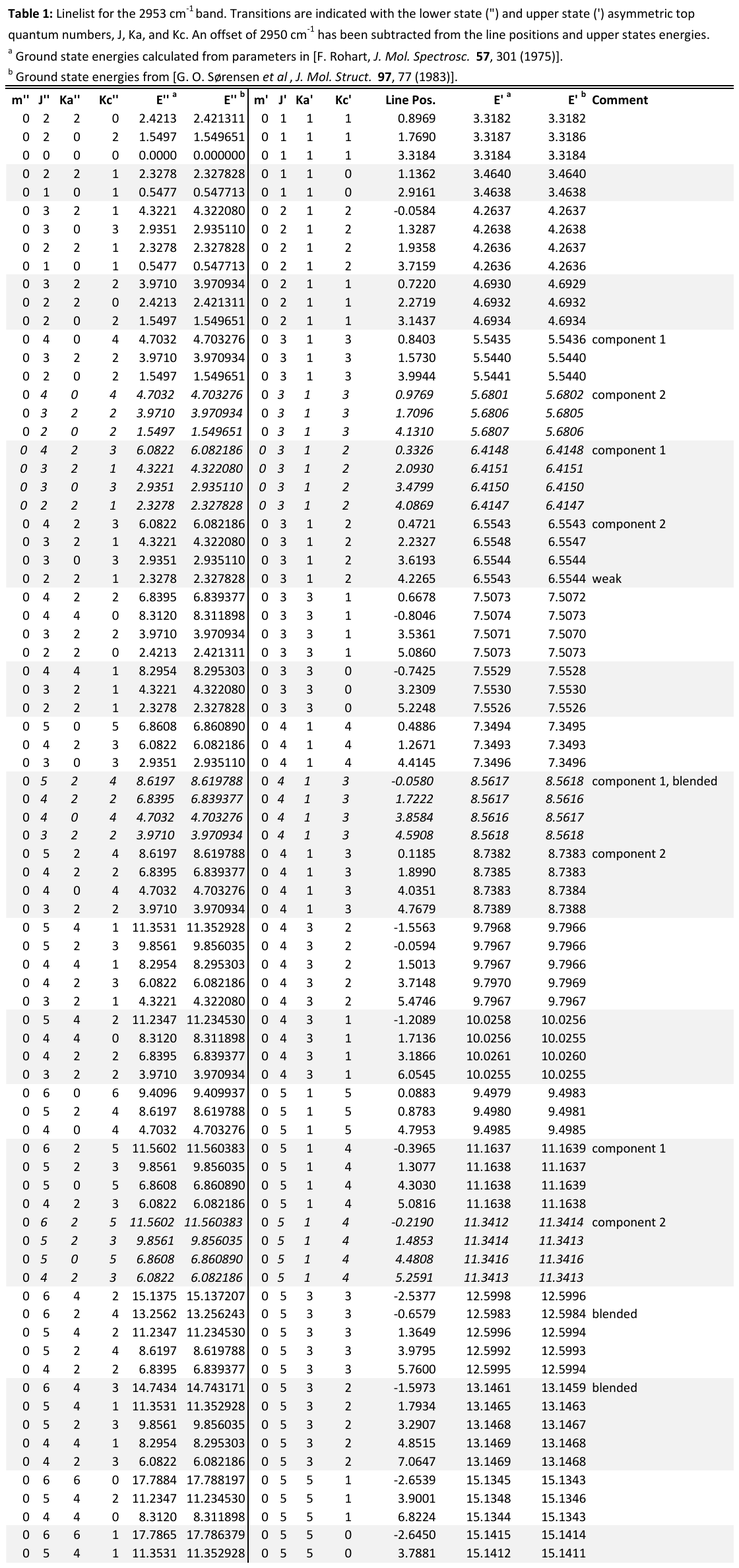}
\includepdf[pages=-]{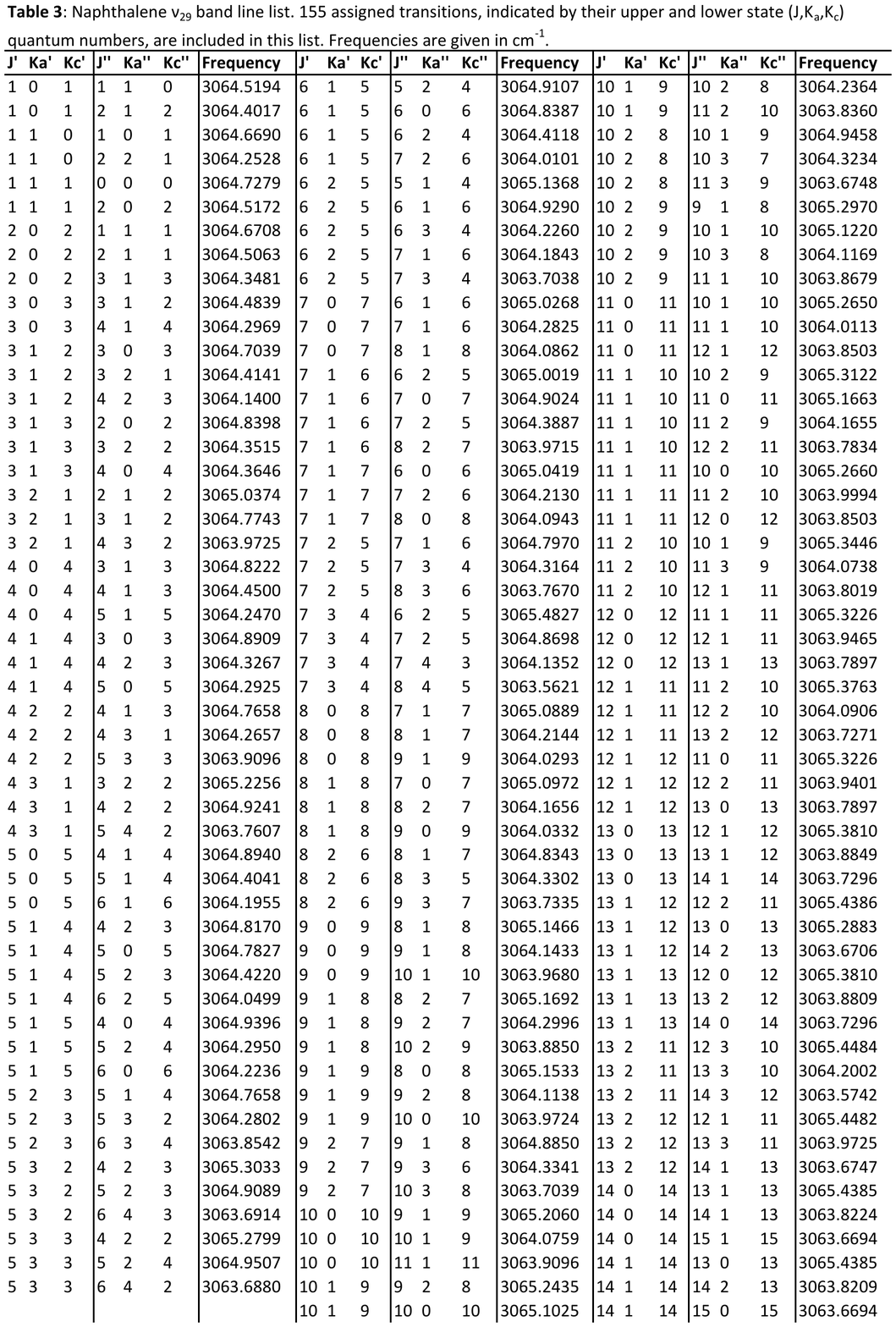}

\end{document}